\documentclass[letterpaper, 10 pt, conference]{ieeeconf}  

\IEEEoverridecommandlockouts                              

\usepackage{amsmath}
\DeclareMathOperator*{\argmax}{arg\,max}

\usepackage{amssymb}  

\usepackage{subcaption}
\usepackage{graphicx}
\usepackage{booktabs}
\usepackage{hyperref}

\title{\LARGE \bf
Adaptive Driving Style for SAE Level-2 Driving Automation: Minimizing Preference Mismatch
}

\author{Kumar Akash$^{1}$, Zhaobo Zheng$^{1}$, Teruhisa Misu$^{1}$, \\Vidya Krishnamoorthy$^{2}$, Mia Dong$^{2}$, Yuni Lee$^{2}$, and Gaojian Huang$^{2}$
\thanks{$^{1}$
        Honda Research Institute USA, Inc., San Jose, CA, USA
        {\tt\small \{kakash, zhaobo\_zheng, tmisu\}@honda-ri.com}}%
\thanks{$^{2}$San Jose State University,
        San Jose, CA, USA
        {\tt\small \{vidya.krishnamoorthy, miaomiao.dong, yuni.lee, gaojian.huang\}@sjsu.edu}}%
\thanks{This is the author’s version of the work. The definitive Version of Record is published in the 2026 American Control Conference (ACC), May 26–29, 2026, New Orleans, Louisiana, USA. Personal use of this material is permitted.  Permission must be obtained for all other uses, in any current or future media, including reprinting/republishing this material for advertising or promotional purposes, creating new collective works, for resale or redistribution to servers or lists, or reuse of any copyrighted component of this work in other works.}%
}

\begin{document}

\maketitle
\thispagestyle{empty}
\pagestyle{empty}

\begin{abstract}

Driving style is a key factor in the comfort and acceptance of automated vehicle (AV) features. In SAE Level-2 automation, where the driver must supervise the system and remain ready to intervene, mismatches between the automation’s driving style and the driver’s preference can reduce trust and trigger takeovers. This paper proposes an adaptive driving-style control framework that minimizes such preference mismatch. In a driving-simulator study, we compare fixed, trust-based, and preference-based adaptation heuristics and analyze their effects on preference mismatch and trust. We then train a driving-preference prediction model and use it in an implicit adaptation policy that selects among bounded driving styles for upcoming events. A validation study shows that the predictive policy achieves equal or lower preference mismatch than comparison baselines, particularly when starting from a less defensive style, while also yielding higher average trust. The results provide a step toward developing human-aware driving automation that can implicitly adapt its driving style to the driver’s preferences.

\end{abstract}

\section{Introduction}

Advances in automated driving systems have enabled automated vehicles (AVs) to operate with increasing levels of autonomy. These technologies have the potential to provide safer and more comfortable transportation while contributing to socially and environmentally sustainable mobility \cite{eugensson2013environmental,gold2015trust,krueger2016preferences}. However, the realization of these benefits depends strongly on user adoption. User acceptance of automated vehicles is closely tied to drivers' trust in the system \cite{adnan2018trust, choi2015investigating}. Beyond reliability and safety, automated vehicles must also account for user comfort and perceived driving quality \cite{hartwich2018driving}. Preferences and perceptions of comfort can vary significantly across users and even within a single user, depending on their mental state or the driving situation \cite{dillen2020keep}. Therefore, systems that can adapt and personalize their behavior to the user have greater potential for widespread acceptance.

A key determinant of perceived comfort in automated driving is the driving style of the automation \cite{dettmann2021comfort, bellem2018comfort}. Drivers differ in their preferences for driving style; some prefer a more defensive approach, characterized by lower speeds and greater distance from surrounding vehicles, while others prefer a more aggressive approach, characterized by higher speeds and shorter gaps. Moreover, a driver's preferred driving style may change dynamically during a drive depending on context, trust, and prior experience \cite{akash2020toward,kleisen2011relationship}. Several studies have examined how different automated driving styles influence user perception. For example, Lee et al. showed that defensive automated driving tends to increase accelerator interventions while aggressive driving increases braking interventions \cite{lee2021assessing}. Basu et al. found that a driver's personal driving style does not always align with their preferred automated driving style, and that users often behave more defensively when interacting with automation \cite{basu2017you}. Bellem et al. demonstrated that changes in acceleration and jerk profiles allow users to distinguish between different automated driving styles \cite{bellem2016objective}. However, most of these studies focus on fixed, non-adaptive driving styles.

Recent research has begun exploring adaptive and personalized driving behavior. Our prior work and related studies have developed several building blocks for adaptive driving-style personalization in Level-2 automation. Sajedinia et al. investigated heuristic adaptive driving styles and found that trust, scenario type, braking behavior, and AV aggressiveness are important correlates of user preference and takeover \cite{sajedinia2022investigating}. Zheng et al. demonstrated that preferred driving style can be inferred from implicit multimodal signals in SAE Level-2 vehicles \cite{zheng2022identification}. Koochaki et al. further extended this line of work with a learnable multimodal preference-prediction framework that can evolve into individualized models as additional user data becomes available \cite{koochaki2023learn}. Complementary behavioral studies have also shown that trust-based adaptive interaction and event type influence trust, preference, and takeover behavior. More broadly, recent work on personalized automated driving, such as MAVERIC, explores data-driven user-specific controller personalization \cite{schrum2024maveric}, highlighting the growing importance of runtime personalization in automated driving systems.

In this paper, we focus on SAE Level-2 driving automation, which corresponds to partial driving automation \cite{SAEInternational}. In Level 2 systems, the automation performs sustained longitudinal and lateral vehicle motion control while the human driver remains responsible for supervision and intervention when necessary. Because the driver remains in the control loop, mismatches between the automation's driving style and the driver's momentary preference can manifest directly as brake or throttle takeovers, reductions in trust, or requests for different driving behavior. This makes Level 2 a particularly relevant setting for studying adaptive driving-style personalization.

Adaptation in advanced driver assistance systems is typically categorized as either explicit or implicit \cite{hasenjager2017personalization}. Explicit adaptation requires the user to directly select system preferences, such as adjusting the following distance in adaptive cruise control. While explicit adaptation provides transparency and user control, it can increase workload and may expose users to undesirable driving behavior before the preferred configuration is reached. Implicit adaptation, in contrast, infers user preferences from observed behavior and adjusts system parameters accordingly. However, implicit approaches require models capable of predicting changes in preferences from contextual and behavioral signals. Prior work on adaptive driving strategies has often focused on specific parameters such as acceleration \cite{kesting2007extending}. In contrast, we investigate event-level preference prediction using dynamic information, including takeover actions, road scene context, and the vehicle's current aggressiveness.

In this work, we develop both explicit and implicit algorithms for adaptive driving-style control to reduce preference mismatch and improve driver trust. First, we evaluate heuristic adaptive strategies that adjust driving style based on explicit trust and preference feedback. We collect data from 36 participants interacting with an SAE Level-2 automated vehicle in an urban driving simulator. Using this dataset, we analyze the relationship between trust, preference change, and takeover behavior. We then develop a predictive model of driving-style preference change and derive an optimal control policy that selects the driving style most likely to minimize preference mismatch. A second validation study with 12 participants demonstrates that the proposed implicit adaptation policy reduces preference mismatch and increases trust compared with heuristic baselines. Although an adaptive driving style is relevant at higher levels of automation, Level 2 provides a unique setting in which a preference mismatch directly translates into observable driver interventions and changes in trust. Therefore, the scope of this work focuses on Level 2 automation.

This paper makes three contributions:
\begin{itemize}
    \item We experimentally compare fixed, trust-based, and preference-based driving-style adaptation policies in a human-subject study of Level-2 automation.
    \item We evaluate whether event-level trust change can serve as a practical proxy for preferred style change and show that this relationship is informative but asymmetric.
    \item We develop and validate an implicit predictive controller that uses event context and takeover behavior to select the driving style most likely to minimize preference mismatch.
\end{itemize}
\section{Human Subject Data Collection}\label{sec:study}

We first conducted a human-subject study to analyze \emph{explicit} adaptation of automated driving styles by comparing heuristic-based adaptive policies with fixed driving styles. The collected data were then used to train a prediction model for preferred changes in AV driving style, which is necessary for developing an \emph{implicit} adaptation policy. 

\subsection{Participants}
\label{sec:participants}
We recruited 36 participants (mean age = 22.86, standard deviation = 5.487, range: 18--38), including 16 females, 19 males, and 1 non-binary participant. All participants met the following requirements: 1) possession of a valid driver’s license, 2) 18 to 65 years of age, and 3) no self-reported sensory deficiency. The entire experiment lasted approximately two hours. The study was approved by the San Jose State University Institutional Review Board (approval ID: 21232).

\subsection{Apparatus}
\label{sec:Apparatus}
\begin{figure}
    \centering
    \includegraphics[width=.85\linewidth]{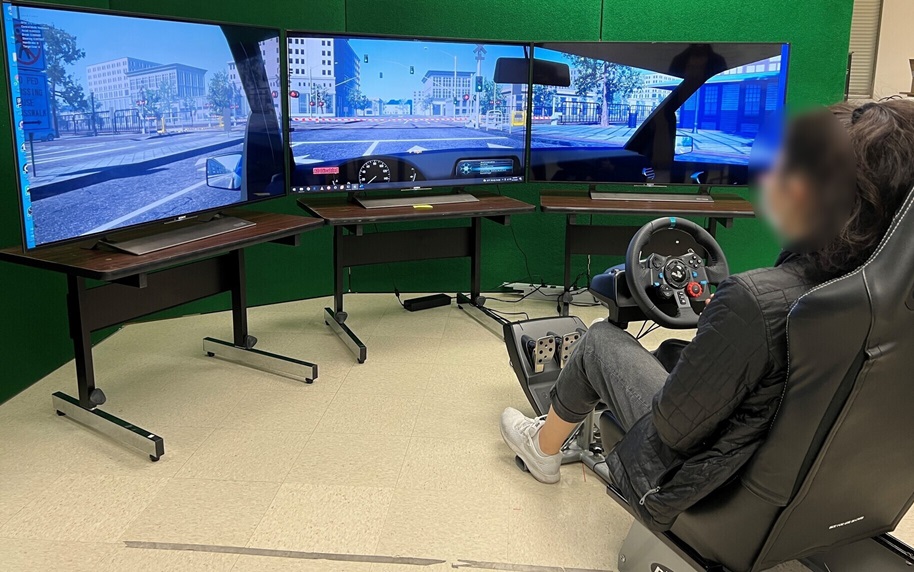}
    \caption{Medium-fidelity driving simulator with three screens, steering wheel, brake pedal, and accelerator pedal.}
    \label{fig:drive_sim}
\end{figure}

Data were collected using a medium-fidelity driving simulator (Figure \ref{fig:drive_sim}) with a steering wheel, brake pedal, and accelerator pedal, and three 45-inch TV screens displaying the vehicle dashboard and surrounding driving environment. The dashboard included a speedometer, a navigation arrow indicating the next target direction, and the driving automation's on/off status. The driving environment was rendered in Unreal Engine \cite{epicgames2019unreal} and simulated a city environment with traffic lights, other vehicles, pedestrians, stop signs, and roundabouts. 

\subsection{Design}
\label{sec:Design}
\begin{figure}
\centering
\begin{subfigure}{0.4\textwidth}
\centering
\includegraphics[width=\linewidth]{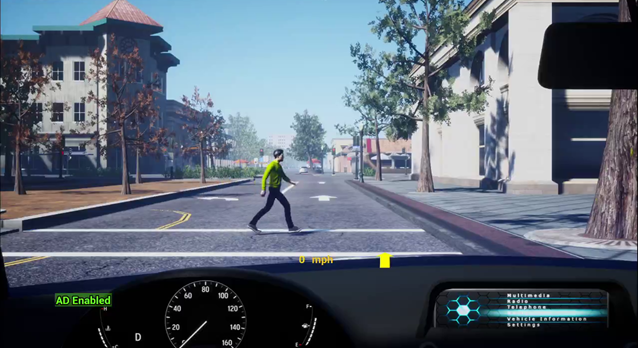} 
\caption{Pedestrian-related event with a pedestrian crossing at the crosswalk.}
\label{fig:ped_crosswalk}
\end{subfigure}
\begin{subfigure}{0.4\textwidth}
\centering
\includegraphics[width=\linewidth]{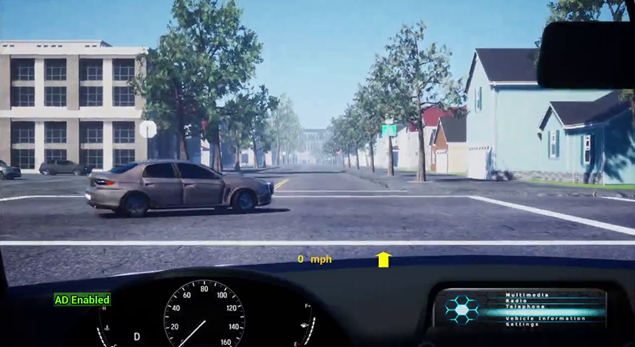}
\caption{Car-related event at a two-way stop.}
\label{fig:2waystop}
\end{subfigure}
\caption{Screenshots showing example events.}
\label{fig:events}
\end{figure}

To understand users' responses to different driving style adaptations, we designed a within-subject study in which each participant completed six automated driving sessions, each with a different driving style adaptation. As shown in Table \ref{tab:def_drive_type}, there were two fixed and four adaptive drive types. Participants were not informed which adaptation policy was active during a given automated session. In each session, the AV passed through sixteen urban intersections, and every other intersection contained an event. Thus, each session included 8 events from two event categories: \emph{pedestrian-related} and \emph{car-related}. Pedestrian-related events included pedestrians on the sidewalk, crossing at the crosswalk, and at the intersection. Car-related events included right turns at a red light, following a lead vehicle, yield and left turns, and a two-way stop. All eight events were presented in each driving session in a randomized order. Figure~\ref{fig:events} shows screenshots of example pedestrian-related and car-related events.

\begin{table*}[]
    \caption{Summary of the six driving style adaptations across the six driving sessions experienced by the participants.}
    \label{tab:def_drive_type}
    \centering
    \begin{tabular}{cl}
    \toprule 
    Name & Driving style adaptation definition \\ \midrule
    $\text{fixed}_\text{LA}$ & Non-adaptive driving style with less aggressive (LA) style throughout all events \\ 
    $\text{fixed}_\text{LD}$ & Non-adaptive driving style with less defensive (LD) style throughout all events \\ 
    $\text{trust}_\text{LA}$ & Adaptive driving style based on the user's trust, starting with a less aggressive style \\ 
    $\text{trust}_\text{LD}$ & Adaptive driving style based on the user's trust, starting with a less defensive style \\ 
    $\text{pref}_\text{LA}$ & Adaptive driving style based on the user's preference, starting with a less aggressive style \\ 
    $\text{pref}_\text{LD}$ & Adaptive driving style based on the user's preference, starting with a less defensive style \\ 
    \bottomrule
    \end{tabular}
\end{table*}

Participants were asked to take over by pressing the throttle or brake pedals whenever they anticipated unsafe or uncomfortable driving. The AV resumed control once the participant removed their input from both the throttle and brake for two continuous seconds. For each event, we annotated the takeover\textsubscript{brake} and takeover\textsubscript{throttle} events, indicating whether the participant pressed the brake or throttle at any point during the event.

The automated driving was implemented using a modified intelligent driver model (IDM) and Stanley controller based on the framework defined in \cite{natarajan2022toward}. We designed four levels of driving style: 1) highly aggressive (HA), 2) less aggressive (LA), 3) less defensive (LD), and 4) highly defensive (HD). These styles vary with driving parameters such as speed, acceleration/deceleration, and the minimum distance to decelerate (MDD), as shown in Table~\ref{tab_IDMparams}. The four styles in this paper represent a bounded, mostly longitudinal aggressiveness-defensiveness axis chosen for interpretability and experimental control rather than as a complete representation of multidimensional human driving style.

\begin{table}
\centering
\caption{Parameters used for the IDM controller for the four driving styles.}\label{tab_IDMparams}
\begin{tabular}{lcccc} 
\hline
{Parameter name} & HA & LA & LD & HD \\ 
\hline
Speed (m/s)             & 14 & 13 & 12 & 11 \\
Max acceleration (m/s$^2$)  & 5 & 4 & 3 & 1 \\
Max deceleration (m/s$^2$)  & 6 & 5 & 2 & 1.5 \\
MDD at intersection (m) & 20 & 15 & 8 & 5 \\
MDD from a pedestrian (m) & 28 & 22 & 15 & 12.5 \\
MDD from a car (m)      & 8 & 9 & 11 & 12 \\
Stop sign duration (s)  & 1.8 & 2 & 2 & 3 \\
\hline
\end{tabular}
\end{table}

We measured both trust and preference because they serve different roles in adaptation. Preference is the direct control target: it indicates whether the current automated driving style should become more aggressive, remain unchanged, or become more defensive. Trust, in contrast, is a low-burden acceptance-related signal that can be queried quickly after each event and therefore offers a practical candidate surrogate for adaptation. Comparing the two lets us test whether trust can substitute for explicit preference feedback in a driving-style adaptation policy.

The two fixed driving style adaptations used LA and LD throughout all events in a driving session. We did not consider HA and HD for fixed driving styles to avoid exposing participants to extreme driving styles throughout an entire drive. For the four adaptive sessions, two used trust-based adaptation, and two used preference-based adaptation, in which the system selected driving aggressiveness from the four levels described above based on the participant's response. These are \emph{explicit adaptations} because they rely on the user's input to change the driving style. Participants were not informed which adaptation policy was active during a given session.

\begin{figure}
\centering
\begin{subfigure}{0.49\textwidth}
\centering
\includegraphics[width=\linewidth]{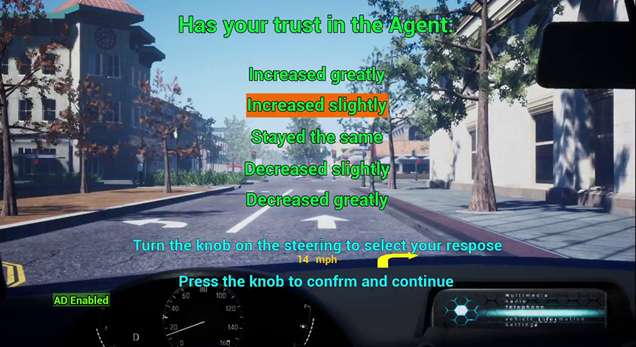}
\caption{On-screen survey to measure trust.}
\label{fig:trust_survey}
\end{subfigure}
\begin{subfigure}{0.49\textwidth}
\centering
\includegraphics[width=\linewidth]{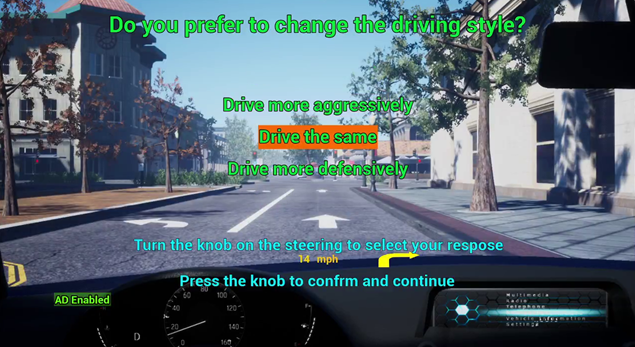}
\caption{On-screen survey to measure preference.}
\label{fig:preference_survey}
\end{subfigure}
\caption{On-screen surveys used to measure trust and preference after each event.}
\label{fig:three}
\end{figure}

In the trust-based adaptive modes, the driving style changed in response to participants' changes in trust in the system. After each event, the one-item trust survey shown in Figure \ref{fig:trust_survey} was presented on the screen. There were five response options: increased greatly (+2), increased slightly (+1), stayed the same (0), decreased slightly (-1), and decreased greatly (-2). After each participant selection, the system recorded the numeric value. When the cumulative change reached +2 or -2, the driving style became more aggressive or more defensive, respectively. For example, two consecutive choices of “decreased slightly” (-1 and -1), or a one-time selection of “decreased greatly” (-2), caused the driving style to become more defensive.

In the preference-based adaptive modes, the driving style changed according to the on-screen preference survey (Figure~\ref{fig:preference_survey}), and each participant's response produced an immediate style update. Specifically, three options were provided: drive more aggressively, stay the same, or drive more defensively. For example, if a participant chose “drive more aggressively” when the vehicle driving style was less aggressive (LA), then the AV driving style moved up one level to highly aggressive (HA). The driving style did not change further once it had reached the limit. For example, the driving style remained HD even when a participant chose “more defensively,” even though it was already HD. We designed the system this way because AVs in real life cannot be excessively defensive or aggressive.

Finally, the fixed mode was used as the baseline, presenting either the LD or LA driving style without adaptation. The trust-based and preference-based adaptive drives each started from either LD or LA as their initial driving style. This resulted in six automated driving style adaptations: $\text{fixed}_\text{LD}$, $\text{fixed}_\text{LA}$, $\text{trust}_\text{LD}$, $\text{trust}_\text{LA}$, $\text{pref}_\text{LD}$, and $\text{pref}_\text{LA}$. Table~\ref{tab:def_drive_type} summarizes these driving style adaptations.

\subsection{Procedure}
\label{sec:Procedure}

Upon arrival, participants signed the consent forms and were briefed on the experimental procedures and the definitions of trust in automation \cite{lee2004trust} and trust in AVs \cite{SAEInternational}. The experiment consisted of seven 10-minute driving sessions: one manual drive and six automated (i.e., SAE Level 2 automation) drives. The manual drive ensured that participants were familiar with the simulator environment and could confidently control the vehicle if they intended to take over the driving automation. There was a 10-minute break between the third and fourth sessions, and the entire experiment lasted approximately two hours.

Participants were asked to keep their dominant hand on the steering wheel and their foot on or near the pedals to standardize takeover readiness during the Level-2 task. To reduce the need to memorize the driving environment and road events, each automated drive session included two separate driving routes. To reduce order effects, the automated drives were presented in a Latin-square counterbalanced order. The manual drive and each automated drive consisted of all eight events (four pedestrian-related and four traffic-related) in a randomized order. During automated drives, the simulator paused after each event to present questions measuring participants’ real-time changes in trust and preferred driving style (Figure \ref{fig:three}).

\section{Data Analysis and Modeling} \label{sec:analysis}

The data collected from the human subject study can be used to train a model to predict the human preference for AV driving style. But first, we analyze the data to evaluate the performance of the explicit driving style adaptations with respect to participants' preference mismatch and trust using their self-reports of trust and preference changes. 

\subsection{Trust as a Proxy for Preferred Style Change}
\begin{figure}
    \centering
    \includegraphics[width=\linewidth]{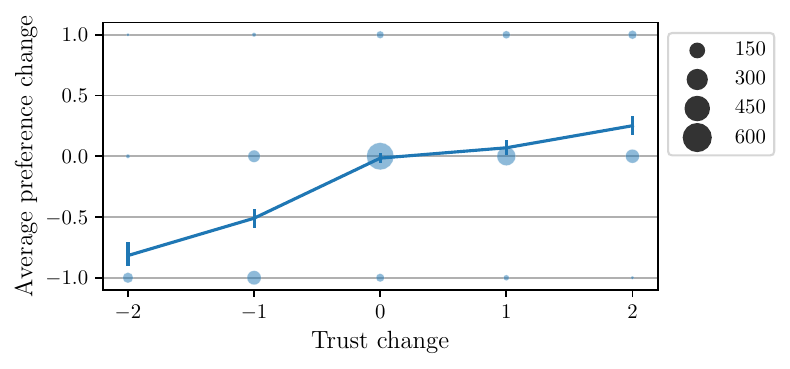}
    \caption{Participants' average change in preference for their change in trust reports. The error bars represent the 95\% confidence interval, and the size of the circles represents the number of events for each preference change and trust change values.}
    \label{fig:trustVsPref}
\end{figure}

The trust-based driving style adaptation heuristic was designed under the engineering approximation that changes in AV driving style preferences are linearly proportional to changes in the user's trust. Specifically, a two-point change in trust will need a one-point change in preference. Trust was measured on a five-level change scale because the heuristic uses accumulated magnitude to trigger a style update, whereas preference was measured on a three-level directional scale because the controller only needs the direction of style adjustment. Figure 4 should therefore be interpreted as an assessment of whether trust can serve as a practical proxy for preferred style change. To examine this approximation, in Fig.~\ref{fig:trustVsPref}, we plot the average change in participants' preference for their reported trust change. Note that the value $+1$ corresponds to `drive more aggressively', $0$ corresponds to `drive the same', and $-1$ corresponds to `drive more defensively'. The results show partial alignment: as trust decreases, larger reductions are associated with a greater desire for defensive driving. However, increases in trust do not consistently imply a desire for more aggressive behavior.

\begin{figure}
    \centering
    \includegraphics[width=1\linewidth]{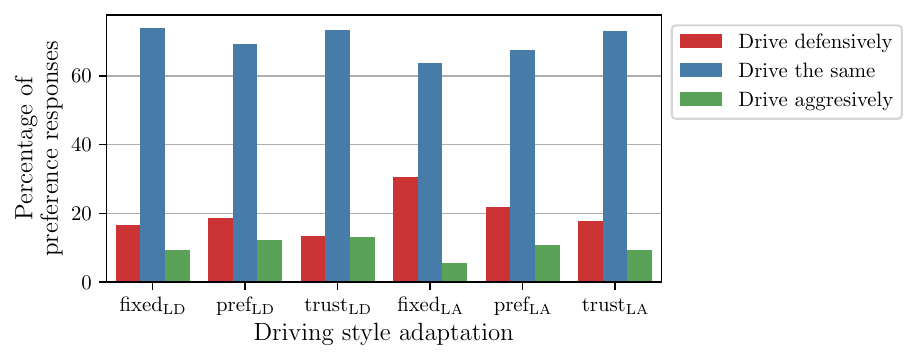}
    \caption{Bar plot representing the percentage of change in preference of participants for each driving style adaptation. Higher `drive the same' percentages indicate lower preference mismatch.}
    \label{fig:pref_exp1}
\end{figure}

Next, we compare self-reported preference changes across the fixed and adaptive drives. Figure~\ref{fig:pref_exp1} shows the percentage of preference change responses for each of the driving style adaptations across all events for all participants. We want to reduce the mismatch between participants' preferred AV driving style and the current AV driving style. Our goal is to achieve higher percentages of `drive the same' responses and thus lower preference mismatches. When comparing the $\text{fixed}_\text{LD}$ with $\text{pref}_\text{LD}$ and $\text{trust}_\text{LD}$ (i.e., the drives that start with less defensive), we see that the trust based heuristic had similar preference mismatch as that of the $\text{fixed}_\text{LD}$. The preference-based heuristic had a relatively higher preference mismatch. A higher `drive the same' percentage indicates lower mismatch, but $\text{fixed}_\text{LD}$ is only one discrete operating point rather than a universally preferred style; because preferred style varies across participants and events, a $\text{fixed}_\text{LD}$ policy can still elicit requests for both more aggressive and more defensive behavior. When comparing the $\text{fixed}_\text{LA}$ with $\text{pref}_\text{LA}$ and $\text{trust}_\text{LA}$ (i.e., the drives that start with less aggressive), we see that the trust based heuristic had lowest preference mismatch followed by preference-based heuristic, and than $\text{fixed}_\text{LA}$. 

\begin{figure}
    \centering
    \includegraphics[width=\linewidth]{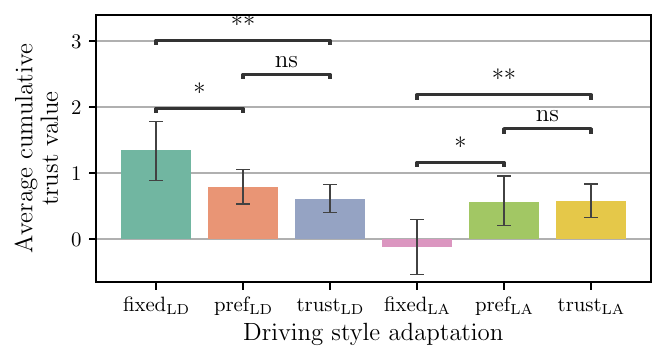}
    \caption{Bar plot representing the average cumulative trust of participants for each driving style adaptation. Note: ns is $\text{not significant}$; *$p<0.05$; **$p<0.01$; ***$p<0.001$.}
    \label{fig:trust_exp1}
\end{figure}

Figure~\ref{fig:trust_exp1} shows the average cumulative trust of the participants across fixed and adaptive drives. The cumulative trust for each event for a participant is calculated by summing all self-report trust change values up to that event. The plot also shows the statistical significance of the comparisons based on two-sample t-tests. When comparing the $\text{fixed}_\text{LD}$ with $\text{pref}_\text{LD}$ and $\text{trust}_\text{LD}$, we see that $\text{fixed}_\text{LD}$ has significantly higher trust as compared to the adaptive drives. However, when comparing the $\text{fixed}_\text{LA}$ with $\text{pref}_\text{LA}$ and $\text{trust}_\text{LA}$, $\text{fixed}_\text{LA}$ has significantly lower trust as compared to the adaptive drives. Both trust- and preference-based adaptive drives have similar trust values for each of LA and LD cases.  

Taken together, Fig.~\ref{fig:pref_exp1} and \ref{fig:trust_exp1} show that trust is informative for adaptation, but not a symmetric surrogate for preferred style change. When the drive starts from the less aggressive condition, the trust-based heuristic yields lower preference mismatch and higher trust than the fixed baseline. When the drive starts from the less defensive condition, trust-based adaptation maintains preference mismatch near the fixed baseline but does not improve trust. A plausible interpretation is that the initial style establishes a reference point against which later behavior is judged, producing an anchoring effect; this may coexist with a general tendency toward moderately defensive preferences in the short urban-simulator sessions. We present this as an interpretation rather than a causal claim because anchoring was not independently manipulated. This asymmetry suggests that the usefulness of trust as an adaptation signal depends on the initial style condition rather than reflecting a universally valid proxy mapping.

This shows that a trust-based heuristic can help reduce preference mismatch and maintain high trust when starting with an aggressive driving style. Moreover, a trust-based heuristic can maintain the same mismatch if starting with defensive, but it will result in lower trust. Surprisingly, the preference-based heuristic performed poorly at reducing preference mismatch. One possible reason is that the preference-based heuristic changes the driving style a posteriori; that is, the driving style is changed for the next event after the participant reports a preference change in the current event. Therefore, if the expected change from the participant is also based on the current situation, the preference change may be `too late'. Therefore, an implicit adaptation method that can predict preferences for the upcoming event can help mitigate this issue.

\section{Modeling Users' Preferred AV Driving Style}\label{sec:model}
\begin{figure}
    \centering
    \includegraphics[width=\linewidth]{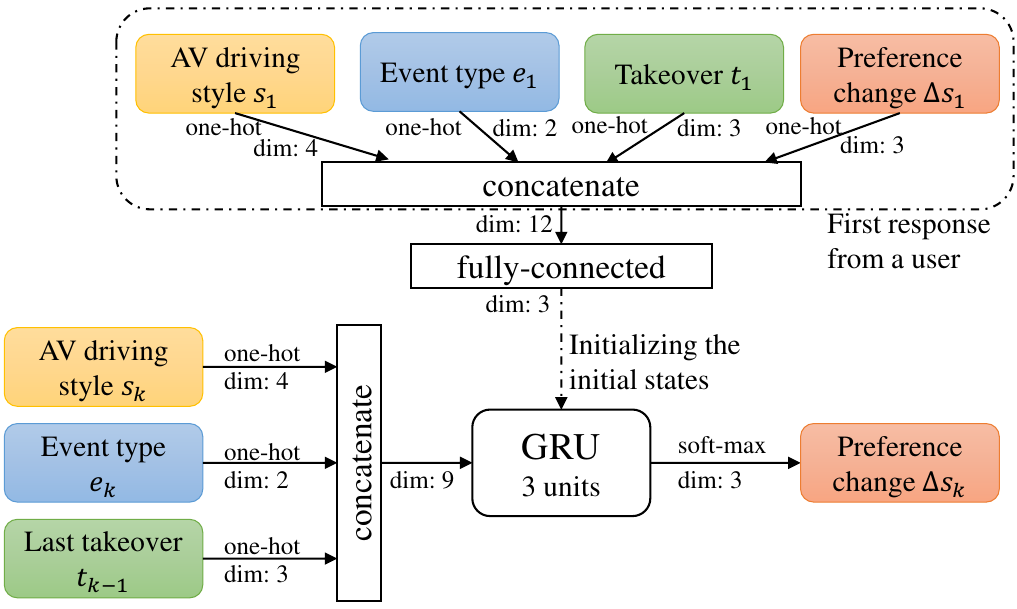}
    \caption{Model structure for preference change prediction.}
    \label{fig:model}
\end{figure}

To develop an implicit driving style adaptation, we first train a model on the collected data to predict users' preferences. We use the AV's current level of driving style (HA, LA, LD, or HD), current event type (car event or pedestrian event), and the user's takeover response in the last event (whether the user pressed brake, pressed throttle, or did not take over) as input to the model. Since the trust and preference behavior is dynamic, we use a gated recurrent unit (GRU) to capture the dynamically evolving behavior\footnote{Other model structures, including CNN, RNN, and LSTM, were also trained, but a comparative study is out of the scope of this paper. The model structure with the best performance is presented.} A relatively small model structure with a limited number of parameters was used to avoid overfitting and to ensure convergence, given the limited data size.  Furthermore, to capture some aspects of individual differences among users, the initial states of the GRU are calculated based on the first situation and each user's corresponding preference change response. Note that only the first preference change response in a drive is used to initialize the model for prediction, and the rest of the responses are only used for analysis later. The model structure is shown in Figure~\ref{fig:model}.

\begin{figure}
    \centering
    \includegraphics[width=\linewidth]{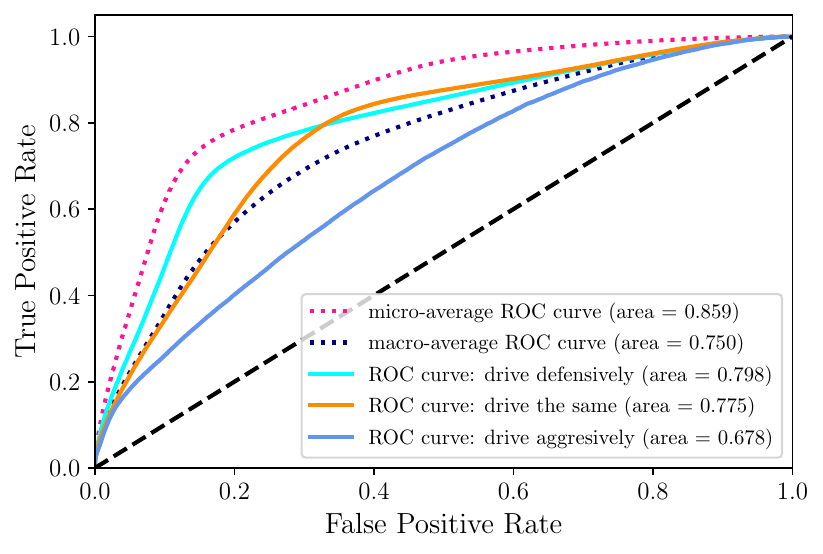}
    \caption{ROC curve and AUC ROC for each class.}
    \label{fig:roc}
\end{figure}

To train the model, we first remove participants with incomplete data, which was mainly due to equipment errors. This results in complete data from 33 participants, with each participant having 6 drives ($33\times6=198$ drives). Each drive comprises a sequence of 8 events. The model is implemented in Keras and trained using the ADAM optimizer \cite{chollet2015keras}. Given the class imbalance across the three preference change classes, we use a weighted loss function with weights inversely proportional to the class frequencies in the training data. We first evaluate the model's performance using 3-fold cross-validation across participants, so that each participant's data appears in only one fold. We repeat the training-evaluation 500 times, with a random 3-fold split each time, to obtain robust metrics. The model is trained for 300 epochs with a batch size of 24. We calculate the accuracy for predicting changes in driving style preference, along with the receiver operating characteristic (ROC) curve and its area under the curve (AUC). Note that, since this is a multiclass prediction (not a binary class), we calculate micro- and macro-averaged ROC curves \cite{vaughan_2022}.  We calculate the mean and the $95\%$ confidence interval for each metric based on 500 iterations. The model structure has an accuracy of $72.12\pm0.13\%$, micro AUC ROC of $0.8585\pm0.0005$, and macro AUC ROC of  $0.7487\pm0.0014$. The ROC curve is shown in Figure~\ref{fig:roc}. Finally, we train the model on the entire dataset for closed-loop control of implicit driving style adaptation. The trained model can continuously predict the probability of a change in driving style preference for the three classes: drive defensively, drive the same, and drive aggressively. Note that the predictor operates over the bounded four-level style axis used in this study, which is primarily longitudinal and gap-related; extending the framework to continuous and multidimensional style adaptation is left for future work. Moreover, more expressive multimodal models can further improve performance and evolve with user-specific data over time.

\section{Control Policy and Validation via User Study} \label{sec:control}

The primary objective of this work is to minimize the mismatch between driving style preferences and changes in both traffic conditions and user preferences. Therefore, an algorithm that maximizes the likelihood of participants responding `drive the same' effectively achieves the objective. For an event number $k \in [1,8]$, let $s_k \in \{\text{HA},\text{LA},\text{LD},\text{HD}\}$ denote the driving style of the AV, $e_k \in \{\text{pedestrian event}, \text{car event}\}$ denote the event type, and $t_k \in \{\text{brake},\text{no takeover},\text{throttle}\}$ denote whether the participant takeover the AV. Therefore, the trained model predicts the likelihood of a participant's preference change $\Delta s_k\in \{\text{drive defensively},\text{drive the same},\text{drive aggressively}\}$ given $s_k$, $e_k$, and $t_{k-1}$, i.e., $p(\Delta s_k|s_k, e_k, t_{k-1})$. Since the event type involves pedestrians and cars, it is fair to assume that it is observable in most cases a priori. Therefore, at an event number $k$ with known event $e_k$, using this model, the best predicted next-event choice of driving style $s^{opt}_k$ under the learned model for the AV is 
$$s^{opt}_k = \argmax_{s_k} \enspace p(\Delta s_k=\text{drive the same}|s_k, e_k, t_{k-1})  \enspace.$$

A control policy that adapts the AV's driving style to the optimal driving style $s^{opt}_k$ can potentially improve users' interaction experience and comfort. We do not optimize for a longer horizon because it is impractical to predict future event types; however, future work can enhance the algorithm by accounting for expected mismatches based on the probability of future events. To validate the control policy, we conducted a human-subjects study similar to the data collection described in Section~\ref{sec:study}. The difference is that we replaced the trust-based driving style adaptations ($\text{trust}_\text{LD}$ and $\text{trust}_\text{LA}$) with the preference change prediction based control policy: $\text{predict}_\text{LD}$ and $\text{predict}_\text{LA}$. Replacing the drives ensured that the total length of the study remained the same and that the counterbalancing order was similar. Similar to the earlier design, $\text{predict}_\text{LD}$ starts with a less defensive style and $\text{predict}_\text{LA}$ starts with a less aggressive style. The participant's first event response is then used to calculate the initial state values for the GRU in the model. Then, for all the upcoming events in the drive, the drive used the control policy to choose the optimal driving style. Although the present controller selects among four bounded driving styles for interpretability and experimental control, the same framework could be extended to continuous adaptation of parameters such as headway, target speed, acceleration/deceleration limits, and gap-acceptance thresholds. In practice, such a controller should include safety bounds, deadbands or hysteresis, and minimum dwell times to prevent oscillatory behavior. We leave this continuous and multidimensional extension to future work.

Twelve participants (mean age = 21.33, standard deviation = 4.25, range: 18--33) were recruited for this study, including 6 females and 6 males. These participants were ensured to be different from those recruited for the first study. All the other screening criteria, compensation, design, and methods were the same as the first study. 

\begin{figure}
    \centering
    \includegraphics[width=1\linewidth]{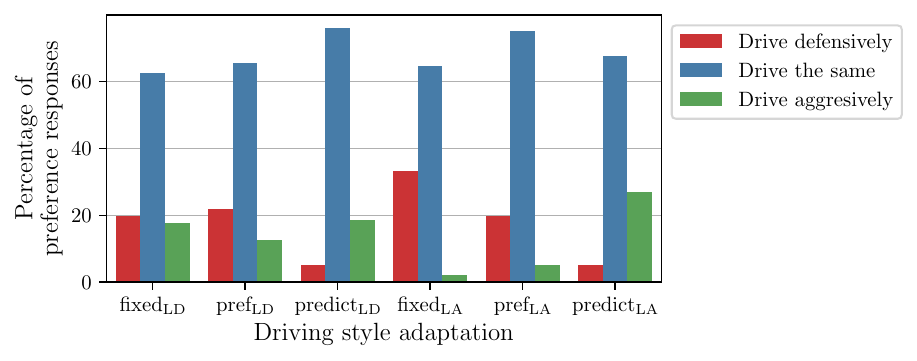}
    \caption{Bar plot representing the percentage of change in preference of participants for each driving style adaptation in the validation study. Higher `drive the same' percentages indicate lower preference mismatch.}
    \label{fig:pref_exp2}
\end{figure}

Now, we compare self-reported preference changes across the fixed and adaptive drives. As the participants in the validation study differ from those in the first study, the values of their self-reports across these two groups may differ due to individual differences; we are only interested in relative within-subject comparisons of driving style adaptations in the validation study. Figure~\ref{fig:pref_exp2} shows the percentage of preference change responses for each of the driving style adaptations across all events for all participants. Note that the higher the percentage of `drive the same' responses, the lower the preference mismatch between participants' preferred AV driving style and the actual AV driving style. When comparing the $\text{predict}_\text{LD}$ with $\text{fixed}_\text{LD}$ and $\text{pref}_\text{LD}$ (i.e., the drives that start with less defensive), we see that the our implicit adaptation control policy had lowest preference mismatch as that of the $\text{fixed}_\text{LD}$ and $\text{pref}_\text{LD}$. However, when comparing the $\text{predict}_\text{LA}$ with $\text{fixed}_\text{LA}$ and $\text{pref}_\text{LA}$ (i.e., the drives that start with less aggressive), we see that the $\text{predict}_\text{LA}$ has lower preference mismatch than $\text{fixed}_\text{LA}$ but higher preference mismatch than $\text{pref}_\text{LA}$. Nonetheless, $\text{predict}_\text{LA}$ achieves this implicitly, without requiring user feedback. Overall, $\text{predict}_\text{LD}$ has the lowest preference mismatch across all driving style adaptations.

\begin{figure}
    \centering
    \includegraphics[width=.9\linewidth]{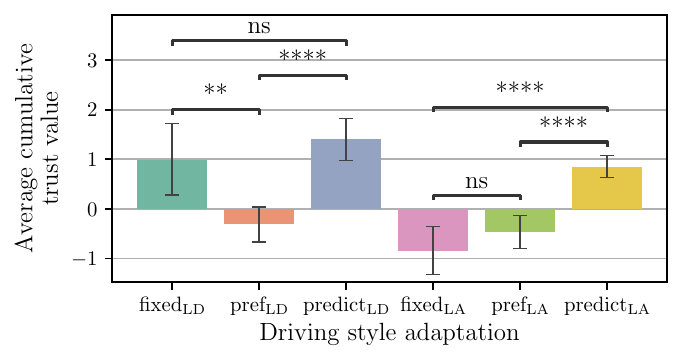}
    \caption{Bar plot representing the average cumulative trust of participants for each driving style adaptation in the validation study. Note: ns is $\text{not significant}$; *$p<0.05$; **$p<0.01$; ***$p<0.001$; ****$p<0.0001$.}
    \label{fig:trust_exp2}\vspace{-20px}
\end{figure}

Figure~\ref{fig:trust_exp2} shows the average cumulative trust of the participants across the fixed and the adaptive drives. Similar to the first study, the cumulative trust for each event for a participant is calculated by summing all self-report trust change values up to that event. The plot also shows the statistical significance of the comparisons based on two-sample t-tests. When comparing the $\text{predict}_\text{LD}$ with $\text{fixed}_\text{LD}$ and $\text{pref}_\text{LD}$, we see that $\text{predict}_\text{LD}$ has significantly higher trust as compared to $\text{pref}_\text{LD}$. Moreover, the average trust of $\text{predict}_\text{LD}$ is higher than $\text{fixed}_\text{LD}$, but is not statistically significant. When comparing the $\text{predict}_\text{LA}$ with $\text{fixed}_\text{LA}$ and $\text{pref}_\text{LA}$, $\text{predict}_\text{LA}$ has significantly higher trust as compared to both $\text{fixed}_\text{LA}$ and $\text{pref}_\text{LA}$, respectively.  

In summary, the results show that our implicit driving-style adaptation policy can reduce preference mismatch, particularly when starting from a less defensive driving style. Moreover, the policy leads to greater trust among participants in the AV. Together, these findings suggest that predictive implicit adaptation can personalize the driving experience without requiring explicit feedback after every event, thereby improving the potential for user comfort and acceptance.

At the same time, several considerations help frame the scope of these results. Because the study was conducted in a medium-fidelity fixed-base simulator, the findings should be interpreted primarily as relative comparisons among adaptation policies under controlled conditions rather than as direct estimates of real-world preferred aggressiveness. Participants were also asked to keep a hand on the steering wheel and a foot near the pedals to standardize takeover readiness during the Level-2 task, which improves experimental control but may reduce ecological naturalness. In addition, the four driving styles represent a bounded, mostly longitudinal aggressiveness--defensiveness axis rather than a complete representation of multidimensional driving style. Finally, because the sessions were relatively short and the participant pool was largely young adults, the observed style preferences may not fully generalize across longer interactions or broader driver populations.

Nevertheless, these results provide encouraging evidence that implicit adaptation is a promising direction for human-aware Level-2 automation. Future work will focus on validating the framework over longer interactions, with broader demographics, and with richer representations of driving style to further improve personalization and trust.

\section{Conclusion} \label{sec:conclusion}
In this paper, we proposed a framework that adaptively changes the driving style of an SAE Level-2 driving automation to match the driver’s preference. We conducted a driving simulator study in which participants interacted with Level-22 driving automation with different heuristic-based driving style adaptations. Results showed that a trust-based heuristic can help reduce preference mismatch and maintain high trust. We then developed a driving preference prediction model to identify the change in preferred driving styles. Using this model, we developed and validated an implicit driving style adaptation algorithm to minimize mismatches in driving style preferences. Results showed that the proposed adaptive algorithm had equal or lower mismatch in preferences and higher average trust in the driving automation. 
This work takes a step toward developing human-aware driving automation that can implicitly adapt its driving style to the driver’s preferences.

\section*{ACKNOWLEDGMENT}
We sincerely acknowledge Kimberly Martinez, Keertana Sureshbabu, and  Brenna Nettles-Miller from San Jose State University for supporting human subject data collection.

\bibliographystyle{IEEEtran}
\bibliography{000_References}             

\end{document}